\documentclass[fleqn,10pt]{olplainarticle}

\usepackage{graphicx}
\usepackage{multicol,multirow}
\usepackage{amsmath,amssymb,amsfonts}
\usepackage{mathrsfs}
\usepackage{amsthm}
\usepackage{rotating}
\usepackage{appendix}
\usepackage{ifpdf}
\usepackage[T1]{fontenc}
\usepackage{newtxtext}
\usepackage{newtxmath}
\usepackage{textcomp}
\usepackage{xcolor}
\usepackage{lipsum}
\usepackage{subcaption}
\usepackage[utf8]{inputenc}
\usepackage[colorlinks,allcolors=blue]{hyperref}

\title{Neural ocean forecasting from sparse satellite-derived observations: a case-study for SSH dynamics and altimetry data}

\author[1]{Daria Botvynko}
\author[1]{Pierre Haslée}
\author[3]{Lucile Gaultier}
\author[2]{Bertrand Chapron}
\author[2]{Clément de Boyer Montégut}
\author[4]{Anass El Aouni}
\author[5]{Julien Le Sommer}
\author[1]{Ronan Fablet}

\affil[1]{Lab-STICC, UMR CNRS 6285, IMT Atlantique, Odyssey, INRIA, IMT Atlantique, Brest, France}

\affil[2]{LOPS, UMR CNRS 6523, Ifremer, Plouzané, France}

\affil[3]{OceanDataLab, Locmaria-Plouzané, France}

\affil[4]{Mercator Ocean International, Toulouse, France}

\affil[5]{CNRS, IRD, Univ. Grenoble Alpes, Grenoble, France}

\keywords{neural forecast, SLA, end-to-end neural scheme, satellite altimetry}

\begin{abstract}
We present an end-to-end deep learning framework for short-term forecasting of global sea surface dynamics based on sparse satellite altimetry data. Building on two state-of-the-art architectures: U-Net and 4DVarNet, originally developed for image segmentation and spatiotemporal interpolation respectively, we adapt the models to forecast the sea level anomaly and sea surface currents over a 7-day horizon using sequences of sparse nadir altimeters observations. The model is trained on data from the GLORYS12 operational ocean reanalysis, with synthetic nadir sampling patterns applied to simulate realistic observational coverage. The forecasting task is formulated as a sequence-to-sequence mapping, with the input comprising partial sea level anomaly (SLA) snapshots and the target being the corresponding future full-field SLA maps.

We evaluate model performance using (i) normalized root mean squared error (nRMSE), (ii) averaged effective resolution, (iii) percentage of correctly predicted velocities magnitudes and angles, and benchmark results against the operational Mercator Ocean forecast product. Results show that end-to-end neural forecasts outperform the baseline across all lead times, with particularly notable improvements in high variability regions. Our framework is developed within the OceanBench benchmarking initiative, promoting reproducibility and standardized evaluation in ocean machine learning. These results demonstrate the feasibility and potential of end-to-end neural forecasting models for operational oceanography, even in data-sparse conditions. 
\end{abstract}

\begin{document}

\flushbottom
\maketitle
\thispagestyle{empty}

\section{Introduction}

The short-term forecasting of the ocean is of key interest for a variety of applications, including among others offshore operations \citep{le2019observation}, maritime traffic routing \citep{davidson2009applications}, ocean extremes \citep{hobday2016hierarchical}, sampling strategies of scientific surveys. State-of-the-art operational systems rely on data assimilation schemes \citep{lellouche2018mercator} to combine a physical model of the ocean dynamics with different observation sources, usually both {\em in situ} and satellite-derived observations. The resulting short-term ocean forecasts typically involve an estimation of the current observation given past available observations. The physical model then propagates the state over the targeted time horizon, typically over a week. Numerous studies support the relevance of the operational ocean forecasting schemes, but also reveal significant uncertainty levels \citep{lellouche2013evaluation}. 

Recently, neural forecasting schemes have emerged as appealing solutions for ocean dynamics, with striking examples for short-term ocean forecasting \citep{wang2024xihe, el2025glonet,cui2025forecasting}. These neural schemes involve emulators of the physical model, trained from reanalysis datasets. These schemes now reach state-of-the-art forecasting performance for short-term ocean forecasts and can even outperform data-assimilation-based systems for specific case-studies and metrics \citep{el2025glonet}.  
While those neural forecasting schemes emerge, the characteristics of ocean observing systems may however question whether one should also address uncertainties in the estimation of initial state by data assimilation schemes. Satellite-derived observations, such as the sea surface height (SSH) and sea surface temperature (SST), involve large sampling gaps, which prevent data assimilation schemes to recover fine-scales dynamics, typically below one or two hundred kilometers for sea surface currents \citep{ballarotta2019resolutions}. The sampling of the interior of the ocean by ARGO floats or moorings is obviously even scarcer. Optimal interpolation has long been the reference approach to deliver gap-free observation-based products, but end-to-end neural mapping schemes \citep{beauchamp2021end, febvre2024training, martin2023synthesizing, martin2024deep} have shown a great potential to better exploit sparse observation datasets and improve the reconstruction of ocean processes.  
 
This study explores how end-to-end neural schemes could contribute to improved short-term ocean forecasts through the direct exploitation of sparse ocean observation datasets. Following the above-mentioned advances in neural interpolation schemes, we state the short-term forecasting of ocean states as an end-to-end neural mapping problem from gappy observations to the targeted ocean forecasts. As demonstration testbed, we focus on SSH forecasts, as SSH is a key signature of global ocean circulation and it is associated with sparse satellite altimetry observations \citep{ducet2000global}. Adapting a state-of-the-art neural mapping scheme \citep{fablet2021learning, febvre2023training}, we do not only adapt the 4DVarNet scheme, but also introduce the state-of-the-art UNet architecture for the forecasting task. Our approach integrates both neural architectures to predict sea surface height from sparse satellite observations. Numerical experiments, conducted with a one-year of real satellite altimetry dataset, demonstrate the potential of these deep learning models to significantly enhance short-term ocean forecasts, compared to the state-of-the-art data-assimilation-based methods. We discuss further these findings towards the evolution of neural ocean forecasting approaches.

This paper is organized as follows. Section \ref{s: problem statement} provides a brief introduction to short-term ocean forecasting.
We present the proposed deep learning approach in Section \ref{s: method}. Section \ref{s: data} describes the considered dataset and benchmark an we report our results in Section \ref{s: results}. Section \ref{s: discussion} discusses our main findings with respect to the state-of-the-art.

\section{Problem statement and Related work}
\label{s: problem statement}

Operational forecasting systems for the earth systems rely on data assimilation schemes \citep{lahoz2014data, buizza2018development}. They state the short-term forecasting of a state of interest as a two-step process. The analysis step solves an initial condition problem and the forecasting comes to propagate the initial condition over the targeted forecasting window using the physical dynamical model. Such assimilation-based systems leverage state-of-the-art general circulation models and data assimilation schemes. GLO12 ocean forecasting system \citep{lellouche2018mercator} falls into this category. It combines NEMO ocean model \citep{madec2015nemo} with a Kalman assimilation scheme \citep{brasseur2006seek} to deliver 10-day global forecasts at 1/12$^\circ$ resolution, providing both hourly and daily averages based on different observation datasets (mainly satellite altimetry, satellite-derived SST and ARGO float data). While representative of the state-of-the-art in terms for forecasting performance \citep{lellouche2013evaluation}, assimilation-based forecasts may involve significant uncertainties especially for time scales below a week and horizontal scales below a few hundreds of kilometers. This may be critical for a variety of applications such as maritime traffic routing \citep{davidson2009applications, lellouche2023evolution}, the forecasting of ocean BGC dynamics \citep{fennel2022ocean}.

From a methodological point of view, assimilation-based system addresses the forecasting of the entire physical state and does not focus on specific variables of interest. Let us introduce the underlying state-space formulation:
\begin{equation}
\label{eq: state space}
\left \{\begin{array}{ccl}
    \displaystyle \frac{\partial \mathbf{x}(t)}{\partial t} &=& {\cal{M}}\left (\mathbf{x}(t) \right )+ \eta(t)\\~\\
    \mathbf{y}(t) &=& {\cal{H}}_t\left ( \mathbf{x}(t) \right ) + \epsilon_m(t), \forall t ,m\\
\end{array}\right.
\end{equation}
where $\mathbf{x}$ is the space-time state of interest and $\mathbf{y}$ is observation process which relates to state $\mathbf{x}$ through observation operator ${\cal{H}}_t$. This observation operator can be time-dependent to account for irregularly-sampled observation data due to the geometry of the observing system (e.g., polar-orbiting earth observation satellites) as well as to the impact of atmosphere conditions onto the passive satellite sensing of the ocean surface (e.g., infrared and multispectral satellite sensors). At a given time $t^*$, the forecasting of state $\mathbf{x}$ from time $t^*$ to time $t^*+N\cdot\Delta$ involves two sub-problems:
\begin{itemize}
    \item the estimation of the state $\widehat{\mathbf{x}}(t^*)$, often stated as a minimisation problem: 
 \begin{equation}
\label{eq: CI}
\widehat{\mathbf{x}}_0 = \arg \min _{\mathbf{x}(t^*)} \mathcal{J} \left ( \mathbf{x}(t^*) , \mathbf{y}(t^*_<) , \mu \right )
\end{equation}
with $\mathcal{J}$ the cost function, $\mathbf{y}(t^*_<)$ the observation time series up to time $t^*$ and $\mu$. Variational data assimilation methods \citep{cummings2013variational} are examples of such schemes.
    \item the iterative application of a time-stepping operator $\Phi_{\mathcal{M},\Delta}$ for dynamical model $\mathcal{M}$ and time step $\delta$ from initial condition $\widehat{\mathbf{x}}_0$ such that:
\begin{equation}
\label{eq: time stepping}
    \mathbf{x}(t^*+n\Delta) = \Phi_{\mathcal{M},\Delta} \left [  \mathbf{x} (t^*+(n-1)\Delta) \right ], \forall n \in [0,N-1]
\end{equation}    
\end{itemize}
This formulation makes explicit two main sources of uncertainties in assimilation-based ocean forecasts:  uncertainties on the estimation of the initial state from available partial observations, and uncertainties associated with the approximation of the true dynamics of state $\mathbf{x}$ by time-stepping operator $\Phi_{\mathcal{M},\Delta}$. The quantification of the these two sources of uncertainties remain challenging \citep{lermusiaux2006quantifying, hoteit2024improving}.

A very recent literature has emerged regarding neural short-term forecast for earth system dynamics \citep{el2025glonet, wang2024xihe, cui2025forecasting}, neural schemes for short-term weather forecasts being the most impactful examples \citep{lam2022graphcast, bi2023accurate}. Interestingly, different neural architectures, including purely data-driven schemes \citep{el2025glonet} and hybrid models using neural correction terms~\citep{liu2023systematic}, have rapidly reached state-of-the-art performance on a global scale compared with model-based forecasts. They typically exploit end-to-end learning strategies from reanalysis datasets and state the short-term forecasting as the prediction of a time series of states over a given time window from a gap-free initial condition. 
The best neural schemes generalize to real-time forecast configurations to compete or outperform operational systems such as ECMWF IFS schemes \citep{bi2023accurate}. Similar learning strategies have also been explored for short-term ocean forecasts \citep{wang2024xihe, xiong2023ai} and points out the potential improvement of assimilation-based ocean forecasts. Most neural forecasting schemes exploit auto-regressive formulations and can be regarded as replacing the model-based time-stepping operator by a neural time-stepping operator:
\begin{equation}
\label{eq: time stepping}
    \mathbf{x}(t^*+n\Delta) = \Phi_{NN} \left [  \mathbf{x} (t^*+(n-1)\Delta) \right ], \forall n \in [0,N-1]
\end{equation}    
As a consequence, such neural forecasting scheme focuses on improving forecasting skills associated to uncertainties in the model-based time-stepping operator $\Phi_{\mathcal{M},\Delta}$.

\begin{figure}[htbp]
    \centering
    \includegraphics[width=\textwidth]{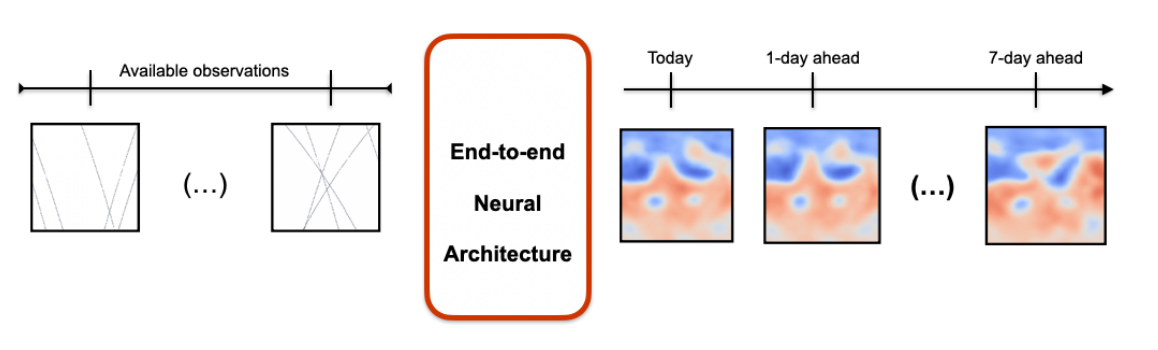}
    \caption{{\bf Proposed end-to-end neural scheme for short-term forecasting:} input satellite observations are fed to the end-to-end neural architecture providing a full SSH forecasted field.}
    \label{fig:end_to_end_neural_forecast}
\end{figure}

Regarding ocean dynamics, the scarcity of the available satellite-derived and in situ observation datasets for the ocean may however question whether the underlying assumption that one can neglect the uncertainties in the estimation of the initial state. Whereas optimal interpolation methods and assimilation-based schemes remain the state-of-the-art operational approaches for most operational oceanography products, numerous studies point out how neural mapping schemes, trained from real observation or simulation-only datasets, could better exploit real gappy datasets for a wide range of missing data rates and patterns, as illustrated among others for satellite altimetry \citep{beauchamp2021end, martin2023synthesizing}, ocean colour products \citep{dorffer2025observation} and sea surface temperatures \citep{barth2020dincae}. These findings invite to question how they translate to short-term ocean forecasts. Through a case-study for satellite altimetry datasets, we explore neural ocean forecasts stated as an end-to-end neural mapping from a series of partial observations to the targeted time series of states in the future, see Figure~\ref{fig:end_to_end_neural_forecast}, that is to say trainable neural operators as:
\begin{equation}
\label{eq: e2e NN}
    \{\mathbf{x}(t^*), \ldots, \mathbf{x}(t^*+N\Delta)\} = \Psi \left [  \mathbf{y} (t^*_<) \right ]
\end{equation}    
In this study we consider the neural operator $\Psi$ to be based on: (i) the neural data assimilation scheme introduced in \citep{fablet2021learning}, or (ii) the state-of-the-art UNet model. In the remainder, we describe those two architectures adapted to the forecast task, as well as the proposed benchmark for the short-term neural forecast of the SSH from satellite altimetry data at global scale.

\section{Proposed approach}
\label{s: method}

\subsection{End-to-end neural schemes for short-term forecasting}

The forecasting problem is formulated as a nonlinear mapping from observed sparse fields at time \( t \) to predicted gap-free fields over the subsequent seven days. Formally:

\[
F: \mathcal{X}_t \rightarrow \mathcal{Y}_{t+1:t+7}
\]

where \( \mathcal{X}_t \) denotes the sparse input and \( \mathcal{Y}_{t+1:t+7} \) the target future sequence. 

Inspired by recent advances in the neural mapping of gappy observation fields \citep{febvre2023training, fablet2023multimodal, martin2023synthesizing}, we address the short-term forecasting problem (\ref{eq: e2e NN}) as the end-to-end learning of a neural parameterization of operator $\Psi$. The underlying statement resorts to considering the short-term forecasting problem as an interpolation problem with observation data available only in the past, {\em i.e.} before time step $t^*$. Specifically, in this study we consider 14-days temporal window in the past.

Considering that the observation data are provided as gridded fields with missing data, any state-of-the-art image-to-image architectures could potentially apply. Among others, we may cite UNet, Transformer and ConvLTSM architectures \citep{ronneberger2015u, ashish2017attention, shi2015convolutional}. As ocean observation data usually involve large missing data rates, neural mapping schemes recently explored for the space-time interpolation of the sea level anomaly (SLA) from satellite altimetry data \citep{fablet2023multimodal, martin2023synthesizing} also appear as appealing solutions.

As described above, in this study we leverage (i) the 4DVarNet architecture, originally designed for spatiotemporal interpolation, and (ii) UNet, the state-of-the-art neural architecture widely used in computational imaging, and adapt them to perform sequence-to-sequence forecasting from incomplete SLA observations. We refer the reader to \citep{fablet2021learning, beauchamp2021end} and to the \hyperlink{https://github.com/CIA-Oceanix/4dvarnet-starter/tree/main}{associated Github page} for the details on the 4DVarNet framework. The UNet code used in this study is an implementation of \citep{ronneberger2015u} and is available in this \hyperlink{https://github.com/milesial/Pytorch-UNet}{Github repository}.

\subsection{OSSE-based training scheme}
\label{ss:osse training}

Following \citep{febvre2023training}, we consider a supervised training scheme from simulation data, while we assess the performance of trained model on real ocean observation data. We detail the considered training dataset in Section \ref{s: data}. Let us denote by $\{\mathbf{x}^{true}_n, \mathbf{y}_n\}_n$ the training dataset formed by pairs of true state and associated gappy observation data, see Figure~\ref{fig:input_and_target}, the latter corresponding to real nadir altimeter sampling patterns. For the 4DVarNet neural scheme the training loss combines the following three elementary losses for each sample index $n$ (while for the UNet it only combines first two):
\begin{equation}
\label{eq: MSE loss}
{\mathcal{L}}_{\mathbf{x},n} = \sum_{k=-K_{O}}^{+K_f} \omega _k \left \| \mathbf{x}^{true}_n(t^*+k\Delta) -  \widehat{\mathbf{x}}_n(t^*+k\Delta) \right \| ^2
\end{equation}
\begin{equation}
\label{eq: grad loss}
{\mathcal{L}}_{\nabla\mathbf{x},n} = \sum_{k=-K_{O}}^{+K_f} \omega _k \left \| \nabla \mathbf{x}^{true}_n(t^*+k\Delta) -  \nabla \widehat{\mathbf{x}}_n(t^*+k\Delta) \right \| ^2
\end{equation}
\begin{equation}
\label{eq: prior loss}
{\mathcal{L}}_{\Phi,n} = \left \| \mathbf{x}^{true}_n - \Phi \left ( \mathbf{x}^{true}_n \right ) \right \| ^2 + \left \| \widehat{\mathbf{x}}_n - \Phi \left ( \widehat{\mathbf{x}}_n \right ) \right \| ^2
\end{equation}
The first two losses refer to MSE losses for state $\mathbf{x}$ and its spatial gradient $\mathbf{x}$, whereas the third loss acts as a regularisation loss \citep{fablet2021learning}. Overall, the training phase amounts to minimizing
a weighted sum of these three losses: $ {\mathcal{L}} = \sum_n\alpha_{\mathbf{x}} {\mathcal{L}}_{\mathbf{x},n} + \alpha_{\nabla \mathbf{x}} {\mathcal{L}}_{\mathbf{x},n} + \alpha_{\Phi} {\mathcal{L}}_{\Phi,n}$. Weights $\alpha_{\mathbf{x}}$, $\alpha_{\nabla\mathbf{x}}$ and $\alpha_{\Phi}$ were tuned empirically to respectively 50, 1000 and 1 for the 4DVarNet and 50 for the UNet.

\subsection{Data Preparation} \label{s: data}
\begin{figure}[htbp]
    \centering
    \includegraphics[width=\textwidth]{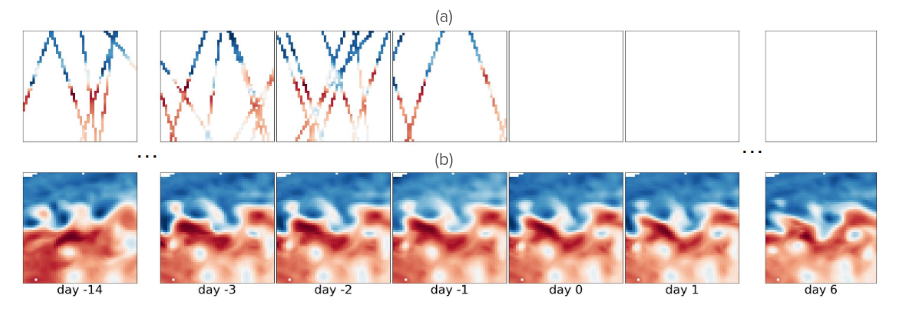}
    \caption{{\bf Input and target of the proposed end-to-end neural scheme for short-term forecasting:} (a) the input to the end-to-end neural model is a set of 14 past days of satellite observations and the corresponding target is (b) a set of 7 future days of complete SSH fields.}
    \label{fig:input_and_target}
\end{figure}

Training is conducted using the GLORYS12 operational ocean reanalysis product, which provides daily global SLA fields at a 1/4$^\circ$ spatial resolution. To simulate real-world altimetric coverage, we generate synthetic nadir-like sampling patterns from the GLORYS12 SLA fields\footnote{Importantly, this SLA is derived from the related SSH field by substracting the corresponding mean dynamic topography (MDT) doi:10.48670/moi-00021} and couple them to the corresponding gap-free fields, see Figure~\ref{fig:input_and_target}. These sampled tracks serve as input to the model, while the corresponding future full-field GLORYS12 outputs are used as ground truth during training. During the inference phase, we use real satellite altimetry data from near-real-time dataset doi:10.48670/moi-00147. 

\subsection{Model Training}

The models were trained over the 2010–2019 period using Adam optimizer and a learning rate e-3. Inputs consist of sequences of spatial SLA fields with missing values due to the nadir sampling mask, and outputs are the corresponding complete SLA sequences over a 7-day forecast horizon. The training process was performed using one A100 GPU and it took a couple of days for the 4DVarNet to converge and around a day for the UNet. We let the reader to our \hyperlink{https://github.com/CIA-Oceanix/4dvarnet-starter/tree/forecast-glorys12}{PyTorch code} for further details.

\subsection{Evaluation Metrics}\label{s: evaluation_metrics}

The performance of the forecasting system is quantitatively evaluated through a set of diagnostics assessing the accuracy and spatial fidelity of the predicted SLA and SSC fields. All metrics are computed by comparing model forecasts against independent observational datasets: SLA from along-track delayed-time Saral/AltiKa dataset doi:10.48670/moi-00146, and velocities of drifters trajectories at 15-m depth doi:10.17882/86236, which were filtered by daily averaging, similarly to \citep{garcia2025orcast}. 

We evaluate the forecasted SSH outputs of: (i) operational ocean forecast GLO12 provided in real time by Copernicus Marine doi:10.48670/moi-00016, (ii) state-of-the-art neural ocean emulators \citep{el2025glonet} and \citep{wang2024xihe} and (iii) the SLA of the studied end-to-end neural ocean forecasts 4DVarNet and UNet.

Following the evaluation protocol of the \hyperlink{https://2024-dc-ssh-mapping-swot-ose.readthedocs.io/}{Ocean Data Challenge for SSH mapping}, we interpolate the forecasted fields onto the observation points. Specifically, for the SLA metrics we interpolate the forecasted data along nadir tracks, while for SSC we perform the interpolation at the drifters locations. In contrast to the SSH mapping evaluation introduced in the mentioned Data Challenge, in this study we adapt the framework to respect a strictly forecast-oriented evaluation workflow. Thus, we only consider one forecasted field per week for each leadtime (where the first leadtime corresponds to Wednesday). As consequence, in order to ensure temporal consistency of the evaluation, for a given leadtime \( \tau \), we consider only observations acquired within a temporal window of \(\pm 0.5 \)~days around the nominal forecast leadtime \( t_0 + \tau \) in the computation of the metrics described bellow. It is important to mention that a particular attention during the evaluation process is put into the choice of MDT \citep{le2017use} to derive SLA from provided SSH outputs and vice-versa. We consider the MDT distributed by Copernicus Marine in order to depict the SSH-derived variables provided by model-based forecasts (GLO12 \citep{lellouche2018mercator}, GloNet \citep{el2025glonet} and XiHe \citep{wang2024xihe}); and the MDT provided in the corresponding altimetry product doi:10.48670/moi-00146. 

We use the normalized root mean squared error (nRMSE) and the average effective resolution to assess the SLA forecast accuracy. The nRMSE quantifies the relative amplitude of model–observation discrepancies, normalized by the observed variability:

\begin{equation}
\text{nRMSE score} = 1 - \frac{\frac{1}{N} \sum_{i=1}^{N} (\hat{y}_i - y_i)^2}{\frac{1}{N} \sum_{i=1}^{N} y_i^2},
\label{eq:nrmse}
\end{equation}

, where $\hat{y}_i$ and $y_i$ represent the predicted and observed data, respectively, and $N$ the total number of observation samples.

The average effective resolution, or $\lambda_x$, is defined as the first wavelength $\lambda$ where the noise to signal ratio (NSR) falls below the chosen threshold $\alpha$ = 0.5 \citep{ballarotta2019resolutions}, see eq.~\ref{eq:lambda_eff}. Here, the NSR is computed as the ratio of the difference between power spectral densities of the forecasted and observed signals to the power spectral density of the observed signal.

\begin{equation}
\lambda_{x} = 
\min \left\{ \lambda \, \big| \, \mathrm{NSR}\leq \alpha \right\}.
\label{eq:lambda_eff}
\end{equation}

In order to obtain the forecasted geostrophic SSC, we differentiate the corresponding SLA forecasts (except the $\pm2^{\circ}$ equator area) and apply drifters-based corrections $MDT_u$ and $MDT_v$ from the MDT product doi:10.48670/moi-00150: 

\begin{equation}
u = - \frac{g}{f} \frac{\partial \text{SLA}}{\partial y} + MDT_u
\end{equation}

\begin{equation}
v = \frac{g}{f} \frac{\partial \text{SLA}}{\partial x} + MDT_v
\end{equation}
\noindent where \( u \) and \( v \) are zonal and meridional components of geostrophic current; parameters $g$ and $f$ represent the gravitational acceleration and the Coriolis parameter. 

Following the SSC evaluation described in \citep{garcia2025orcast}, we use here two metrics: percentage of correctly predicted velocity directions (\( P_{\theta} \)) and percentage of correctly predicted velocity magnitudes (\( P_{|u|} \)), both compared to the drifters velocities at 15-m depth. \( P_{\theta} \) quantifies the proportion of locations where the angular deviation between forecasted and observed velocities is within a tolerance angle \( \Delta \theta_{\mathrm{max}} \) fixed at \( 45^\circ \), see eq.~\ref{eq:angle} and \( P_{|u|} \) measures the proportion of points where the relative error in speed remains below a threshold \( \Delta_{|u|,\mathrm{max}} \) fixed at 25\%, see eq.~\ref{eq:magnitude}.

\begin{equation}
P_{\theta} = 
\frac{1}{N}
\sum_{i=1}^{N}
\mathbb{I}\!\left( |\hat{\theta_i} - \theta_i| < \Delta \theta_{\mathrm{max}} \right)
\times 100,
\label{eq:angle}
\end{equation}

\begin{equation}
P_{|u|} = 
\frac{1}{N}
\sum_{i=1}^{N}
\mathbb{I}\!\left(
\frac{|\,|\hat{u_i}| - |u_i|\,|}{|u_i|}
< \Delta_{|u|,\mathrm{max}}
\right)
\times 100.
\label{eq:magnitude}
\end{equation}\noindent where \( N \) is the total number of collocated observation samples, and \( \mathbb{I}(\cdot) \) denotes the indicator function.

\section{Results}\label{s: results}

In this section, we present the results of the evaluation of the benchmarked forecast models, see Section \ref{sec:benchmarked_models}. We focus on the performance of the presented forecasts in comparison to the Mercator Ocean operational forecast. The evaluation is performed globally, followed by an analysis of regime-dependent performances across different dynamical regimes.

\subsection{Benchmarked models}\label{sec:benchmarked_models}
In this study, we benchmark state-of-the-art neural ocean emulators \citep{el2025glonet, wang2024xihe} and end-to-end neural ocean forecast models against the operational baseline GLO12 \citep{lellouche2018mercator, lellouche2023evolution}. The models considered are summarized in Table~\ref{tab:benchmarked_models}. They are categorized into two groups: neural ocean emulators, which aim to propagate in time the initial state provided by assimilation-based products, and end-to-end neural forecasts, which directly predict the SLA from sparse inputs without relying on any physical model.
We evaluate all models at leadtimes ranging from 0 to 7 days, with their predictions compared against independent altimeter and drifter observations. The benchmark focuses on their ability to reproduce SLA and SSC on the set of the evaluation metrics presented in Section \ref{s: evaluation_metrics}.
\begin{table}[h]
    \centering
    \caption{Benchmarked models and their specifications.}
    \label{tab:benchmarked_models}
    \begin{tabular}{cccc}
        \hline
        \textbf{Model} & \textbf{Type} & \textbf{Resolution}  & \textbf{Nb of parameters}\\
        \hline
        GLO12  & Operational baseline & 1/12$^\circ$ daily &  - \\
        GloNet & Neural emulator & 1/12$^\circ$ daily & 200M \\
        XiHe & Neural emulator & 1/12$^\circ$ daily & 80M \\
        4DVarNet & End-to-end neural forecast & 1/4$^\circ$ daily & 652K \\
        UNet & End-to-end neural forecast & 1/4$^\circ$ daily & 17M \\
        \hline
    \end{tabular}
\end{table}
\subsection{Global-scale performance}
\begin{figure*}[h!]
    \centering
    \includegraphics[width = \textwidth]{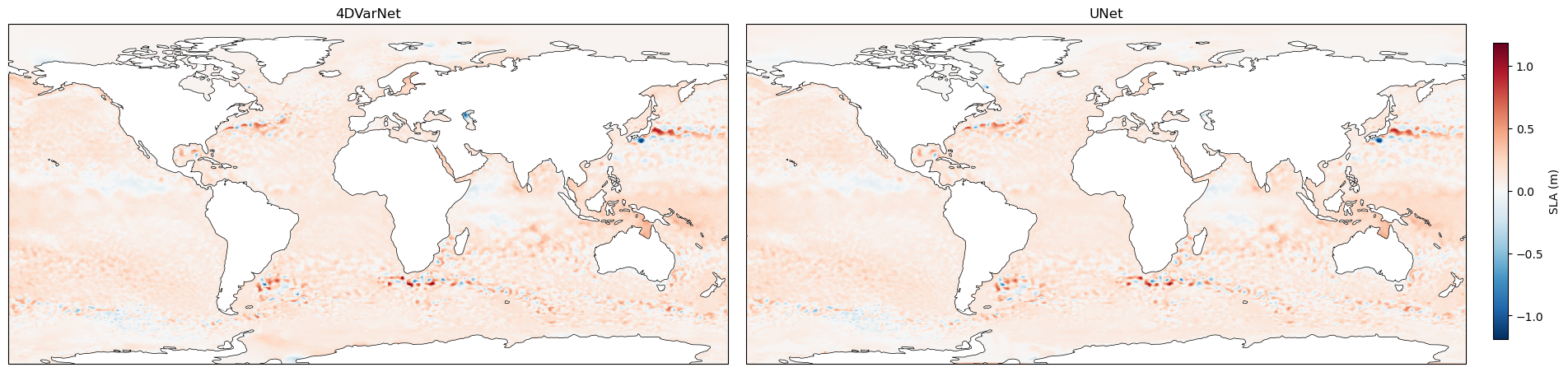}
    \caption{SLA predicted by proposed end-to-end neural forecasts 4DVarNet (left column) and UNet (right column) at leadtime 0 on 2023-01-18.}
    \label{fig:sla_forecasted}
\end{figure*}

\begin{figure*}[h!]
    \centering
    \includegraphics[width=\textwidth]{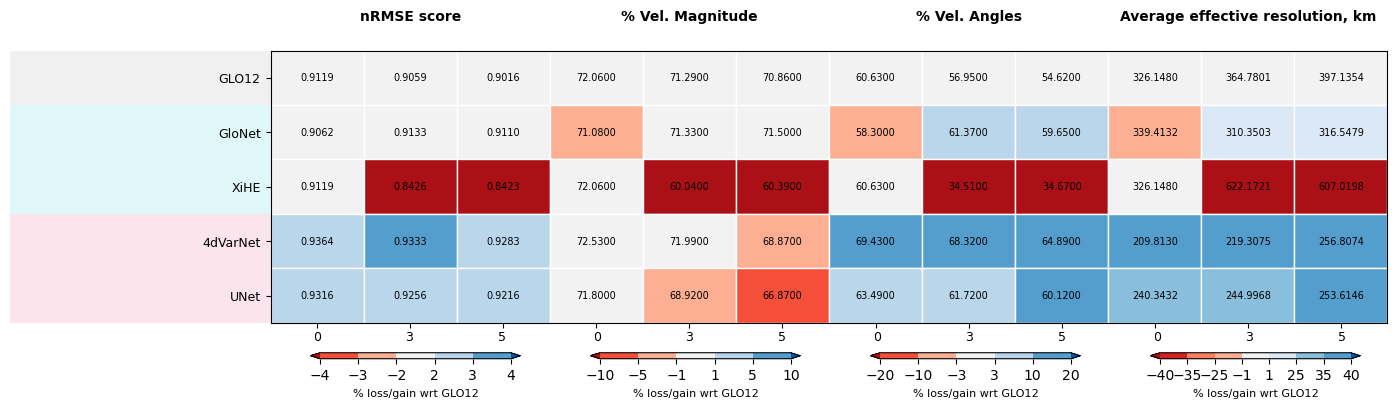}
    \caption{\textbf{Benchmark of SSH forecasts (and derived SSC), provided by GLO12, GloNet, XiHe and proposed end-to-end neural forecasts 4DVarNet and UNet.} The benchmark presents the following metrics: nRMSE score and average effective resolution of the SSH with respect to SARAL/AltiKa, \% correctly predicted SSC angles and magnitudes with respect to drifters. Colors indicate relative gain in \% compared to the baseline GLO12, with blue indicating better performances.}
    \label{fig:global_benchmark}
\end{figure*}

Figure~\ref{fig:sla_forecasted} shows the SLA predicted by end-to-end neural ocean forecasts at leadtime 1 on January 18 2023. Figure~\ref{fig:global_benchmark} shows the evaluation metrics across leadtimes 0, 3 and 5 for the benhmarked neural emulators models, all compared to the baseline GLO12, with the colorbar corresponding to the respective relative gains in \%. The 4DVarNet consistently outperforms baseline GLO12 and other state-of-the-art neural forecasts across the nRMSE SSH score. Performance remains robust over time, highlighting the stability of the model across diverse ocean regimes. Furthermore, 4DVarNet also improves the derived SSC, particularly improving the forecast of SSC angles compared to drifters velocity vectors, suggesting a more accurate capture of flow directions.

\subsection{Regime-dependent performance}
Tables~\ref{tab:relative_error_leadtime0_regions}–\ref{tab:relative_error_leadtime5_regions} assess the performance of the benchmarked models across leadtimes 0, 3 and 5 and three oceanic regimes: coastal, offshore high variability, and offshore low variability regions. Here we inter-compare  the benchmarked models against the observation‑based delayed-time product DUACS \citep{taburet2019duacs}. DUACS uses an Optimal Interpolation (OI) algorithm applied to sparse altimetry data \citep{ballarotta2019resolutions, le2025satellite}. We emphasize that, for a given leadtime, DUACS product exploits past and future altimetry observations. This is the reason why we do not include it in the benchmark among the tested approaches. It provides us with a means to compute relative performance metrics in our regime-dependent analysis. DUACS can better reflect the spatial structures of SLA compared to assimilation-based model analyses, especially in highly dynamic regions \citep{ballarotta2019resolutions}. 
\begin{figure*}[h!]
    \centering
    \includegraphics[width=0.9\textwidth]{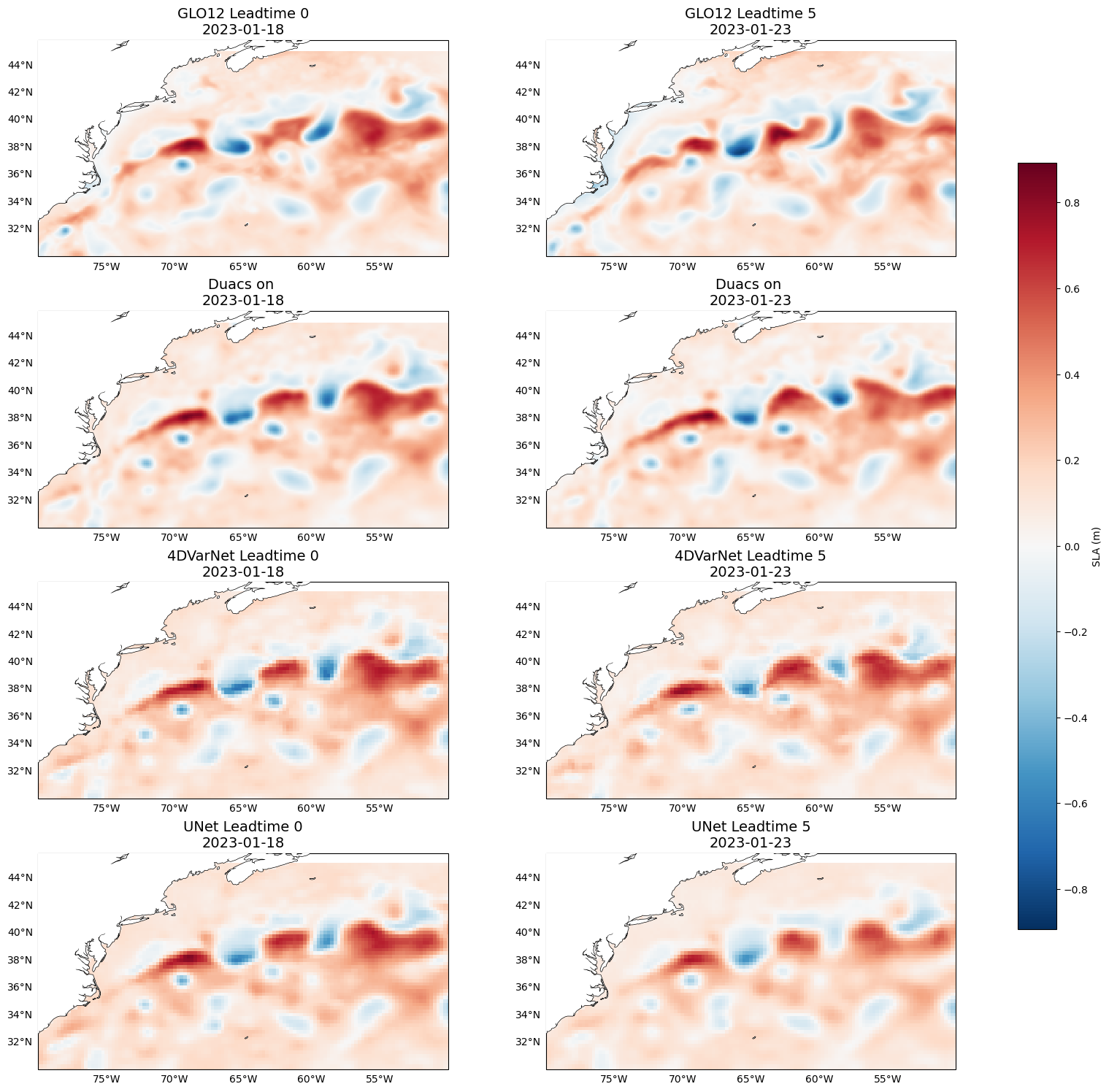}
    \caption{SLA reconstructed by OI-based DUACS product vs baseline forecast GLO12 and proposed end-to-end neural forecasts 4DVarNet and UNet at leadtimes 0 and 5 in the Gulf Stream on 2023-01-18 and 2023-01-23.}
    \label{fig:gulfstream_duacs_glo12}
\end{figure*}

\begin{figure*}[h!]
    \centering
    \includegraphics[width=0.9\textwidth]{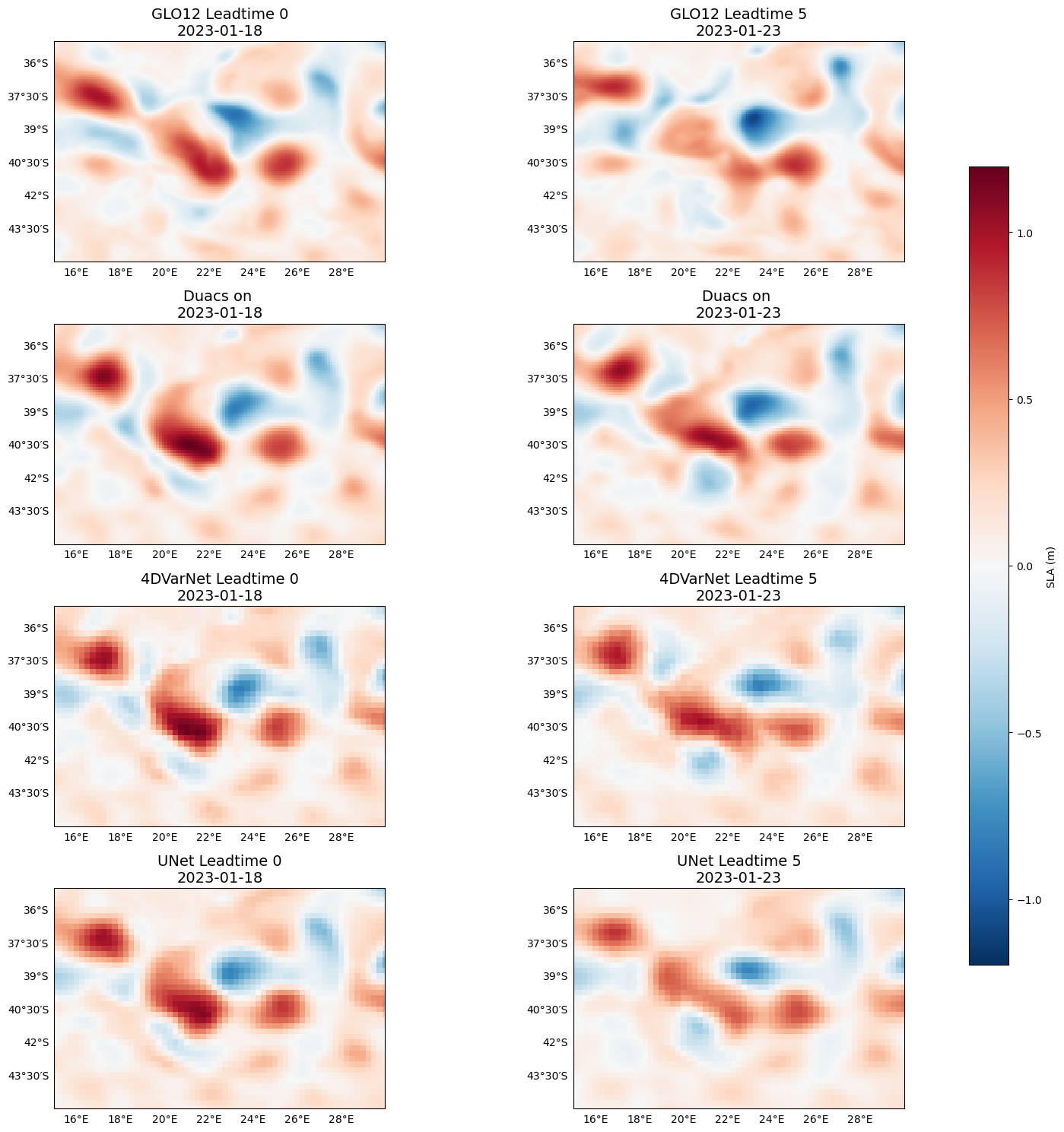}
    \caption{SLA reconstructed by OI-based DUACS product vs baseline forecast GLO12 and proposed end-to-end neural forecasts 4DVarNet and UNet at leadtimes 0 and 5 in the Agulhas region on 2023-01-18 and 2023-01-23.}
    \label{fig:agulhas_duacs_glo12}
\end{figure*}
To quantify forecasts' accuracies relative to the DUACS baseline, we compute percentage gains of each metric relative to the corresponding DUACS score (both evaluated with respect to Saral/AltiKa observations), see Tables~\ref{tab:relative_error_leadtime0_regions}-\ref{tab:relative_error_leadtime5_regions}. At leadtime 0, 4DVarNet achieves the best performance compared to GLO12 baseline across all regions. It remains the closest to the DUACS reconstruction, and is even able to outperform it in offshore high-variability regions. UNet performs slightly worse than 4DVarNet, however outperforms GLO12, especially in coastal regions. GloNet  underperforms 4DVarNet and UNet, while XiHe is at the same level of performance as GLO12, as its first leadtime is defined as the one provided by the reanalysis. At leadtime 3, 4DVarNet maintains strong performance across all regimes, consistently achieving the lowest RMSE and highest nRMSE, reaching positive gains relative to DUACS. UNet shows competitive performance in low-variability areas, but is worse than GloNet in coastal regions, which gains in performance compared to the first leadtime. By leadtime 5, 4DVarNet still consistently outperforms GLO12 across all regimes and is considerably outperforming DUACS reconstruction especially in coastal and offshore low-variance regions. UNet remains competitive especially in low-variance areas. Overall, all models lose performance relative to their first leadtime as forecasts advance, yet 4DVarNet keeping the lowest level of relative errors wrt DUACS, highlighting the ability of this end-to-end neural architecture to remain close to an observations-based short-term forecast even at longer leadtimes.
Figures~\ref{fig:agulhas_duacs_glo12} and \ref{fig:gulfstream_duacs_glo12} further illustrates this comparison by displaying the SLA fields from GLO12, proposed end-to-end forecasts 4DVarNet and UNet at leadtimes 0 and 5, but also DUACS L4 product for the corresponding dates, in two high variability regions: Gulf Stream and Agulhas. As expected, GLO12 struggles to correctly position some dynamical structures in comparison to observation-based product DUACS. In the contrast, the end-to-end neural forecasts better recover spatial patterns and are more consistent with DUACS. At leadtime 5, however, the UNet forecast shows a noticeable loss of energy, reflected in a reduced SLA variance, a phenomenon which is not observed in the 4DVarNet forecasts.
\begin{table*}[h!]
\centering
\caption{Relative error (\%) of RMSE and nRMSE metrics of the forecasted SLA at leadtime 0 with respect to DUACS mapping product. The metrics are calculated for three dynamical regions: coastal, offshore high variability and low variability regions. Positive values indicate degradation relative to DUACS, negative values indicate improvement. Compared to the benchmarked forecasts, DUACS SLA relies on past and future altimetry observations for a given leadtime.}
\label{tab:relative_error_leadtime0_regions}
\begin{tabular}{lcccccc}
\toprule
 & \multicolumn{2}{c}{\textbf{Coastal}} &
   \multicolumn{2}{c}{\textbf{Offshore High Var.}} &
   \multicolumn{2}{c}{\textbf{Offshore Low Var.}} \\
\cmidrule(lr){2-3}\cmidrule(lr){4-5}\cmidrule(lr){6-7}
\textbf{Model} & RMSE (\%) & nRMSE (\%) & RMSE (\%) & nRMSE (\%) & RMSE (\%) & nRMSE (\%) \\
\midrule
GLO12      & 44.81 & 3.13 & 41.52 & 5.91 & 47.22 & 2.50 \\
GloNet     & 52.26 & 3.64 & 52.54 & 7.48 & 57.33 & 3.04 \\
XiHe       & 44.81 & 3.13 & 41.52 & 5.91 & 47.22 & 2.50 \\
\textbf{4DVarNet} & \textbf{8.85} &  \textbf{0.62} &  \textbf{-1.35} &  \textbf{-0.19} &  \textbf{5.07} &  \textbf{0.27} \\
UNet       & 9.97 & 0.70 & 8.95 & 1.27 & 15.91 & 0.84 \\
\bottomrule
\end{tabular}
\end{table*}

\begin{table*}[h!]
\centering
\caption{Relative error (\%) of RMSE and nRMSE metrics of the forecasted SLA  at leadtime 3 with respect to DUACS at the same leadtime. The metrics are calculated for three dynamical regions: coastal, offshore high variability and low variability regions. Positive values indicate degradation relative to DUACS, negative values indicate improvement.}
\label{tab:relative_error_leadtime3_regions}
\begin{tabular}{lcccccc}
\toprule
 & \multicolumn{2}{c}{\textbf{Coastal}} &
   \multicolumn{2}{c}{\textbf{Offshore High Var.}} &
   \multicolumn{2}{c}{\textbf{Offshore Low Var.}} \\
\cmidrule(lr){2-3}\cmidrule(lr){4-5}\cmidrule(lr){6-7}
\textbf{Model} & RMSE (\%) & nRMSE (\%) & RMSE (\%) & nRMSE (\%) & RMSE (\%) & nRMSE (\%) \\
\midrule
GLO12      & 49.23 & 3.52 & 54.25 & 7.72 & 55.51 & 2.97 \\
GloNet     & 39.14 & 2.80 & 34.12 & 4.88 & 46.49 & 2.48 \\
XiHe       & 130.18 & 9.32 & 183.34 & 26.14 & 157.44 & 8.41 \\
\textbf{4DVarNet} & \textbf{11.20} & \textbf{0.80} & \textbf{5.37} & \textbf{-0.77} & \textbf{9.14} & \textbf{0.49} \\
UNet       & 20.47 & 1.46 & 19.49 & 2.78 & 22.79 & 1.21 \\
\bottomrule
\end{tabular}
\end{table*}

\begin{table*}[h!]
\centering
\caption{Relative error (\%) of RMSE and nRMSE metrics of the forecasted SLA  at leadtime 5 with respect to DUACS at the same leadtime. The metrics are calculated for three dynamical regions: coastal, offshore high variability and low variability regions. Positive values indicate degradation relative to DUACS, negative values indicate improvement.}
\label{tab:relative_error_leadtime5_regions}
\begin{tabular}{lcccccc}
\toprule
 & \multicolumn{2}{c}{\textbf{Coastal}} &
   \multicolumn{2}{c}{\textbf{Offshore High Var.}} &
   \multicolumn{2}{c}{\textbf{Offshore Low Var.}} \\
\cmidrule(lr){2-3}\cmidrule(lr){4-5}\cmidrule(lr){6-7}
\textbf{Model} & RMSE (\%) & nRMSE (\%) & RMSE (\%) & nRMSE (\%) & RMSE (\%) & nRMSE (\%) \\
\midrule
GLO12      & 55.21 & 3.85 & 64.23 & 9.39 & 63.32 & 3.35 \\
GloNet     & 48.19 & 3.37 & 35.46 & 5.21 & 50.37 & 2.67 \\
XiHe       & 131.97 & 9.20 & 182.31 & 26.71 & 160.08 & 8.64 \\
\textbf{4DVarNet} & \textbf{15.48} & \textbf{1.08} & \textbf{16.52} & \textbf{2.42} & \textbf{17.72} & \textbf{0.94} \\
UNet       & 26.33 & 1.83 & 29.02 & 4.27 & 29.49 & 1.56 \\
\bottomrule
\end{tabular}
\end{table*}

\begin{figure*}[h!]
    \centering
    \begin{subfigure}[t]{\textwidth}
        \centering
        \caption{}
        \includegraphics[width=\textwidth]{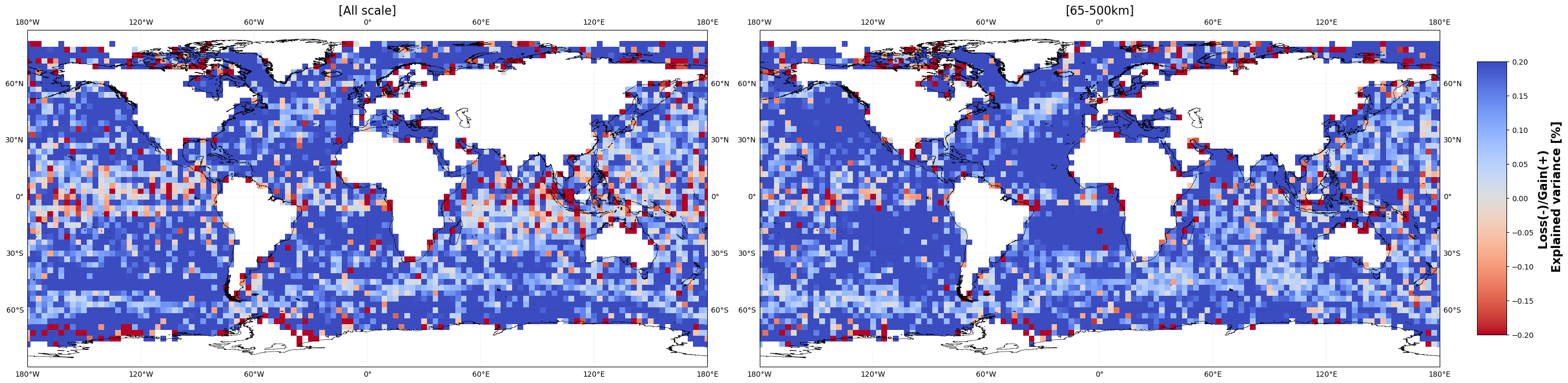}
        \label{fig:subfig1}
    \end{subfigure}
    \vspace{0.8em}
    \begin{subfigure}[t]{\textwidth}
        \centering
        \caption{}
        \includegraphics[width=\textwidth]{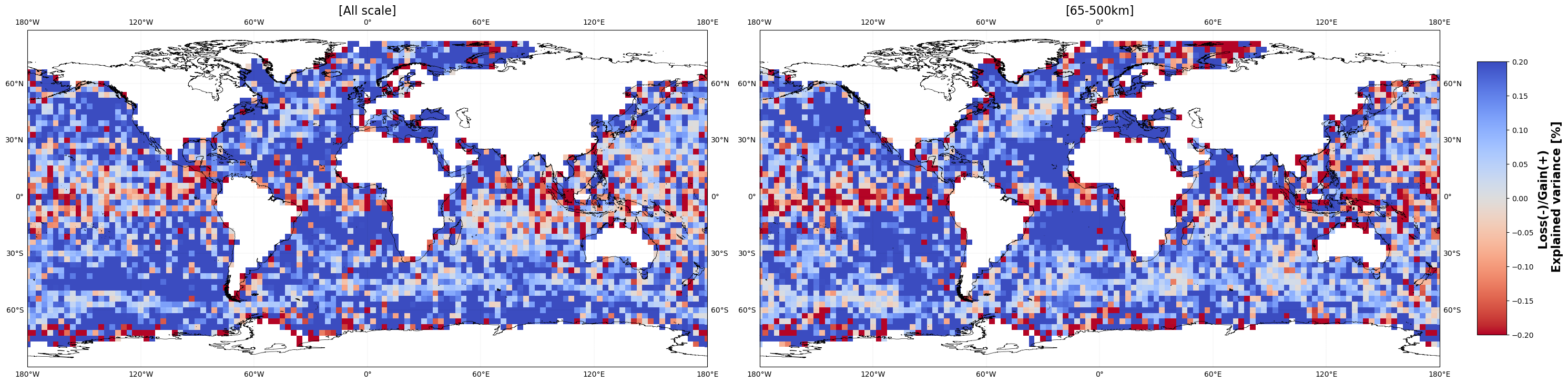}
        \label{fig:subfig2}
    \end{subfigure}
    \vspace{0.8em}
    \begin{subfigure}[t]{\textwidth}
        \centering
        \caption{}        
        \includegraphics[width=\textwidth]{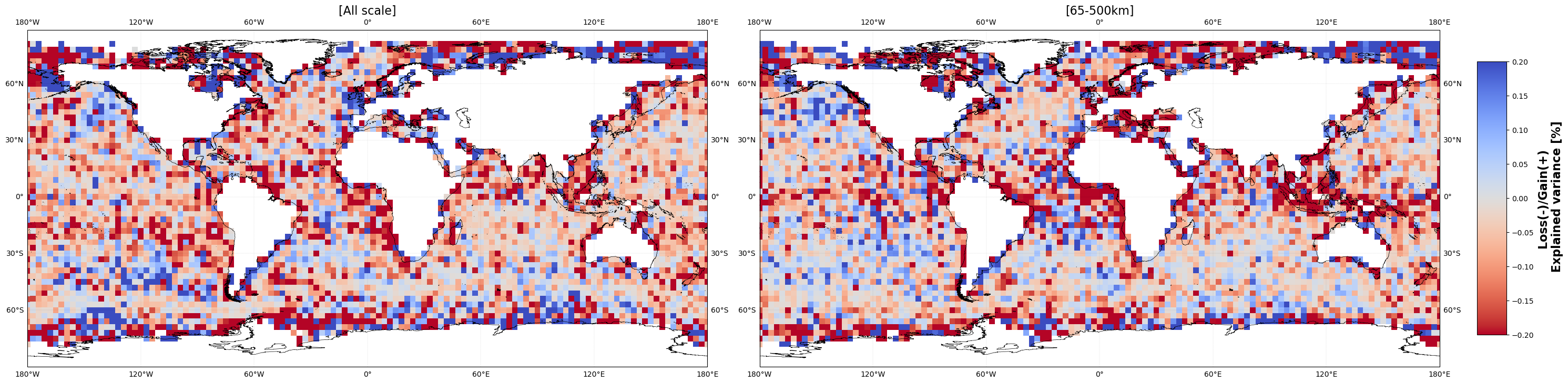}
        \label{fig:subfig3}
    \end{subfigure}

    \caption{\textbf{Gain of the explained variance of the forecasted SLA wrt SARAL/AltiKa observations relative to the one GLO12 baseline at leadtime 0: at all scales (left column) and filtered at 65-500km scales (right column).} The benchmarked forecasts: (a) 4DVarNet, (b) UNet, (c) GloNet.}
    \label{fig:relative_gain_explained_variance}
\end{figure*}

Figure~\ref{fig:relative_gain_explained_variance} presents the relative gains of the benchmarked models with respect to the operational baseline forecast GLO12 for leadtime 0 for the explained variance metric calculated per $1^{\circ} \times 1^{\circ}$ bins. We refer the reader to the \hyperlink{https://2023a-ssh-mapping-ose.readthedocs.io/en/latest/}{SSH mapping data challenge documentation} for further details on the implementation of the metric. The gains are computed as the percentage difference relative to GLO12 for all scales as well as for spatial scales ranging between 65 and 500 km. Visually, the 4DVarNet model shows large improvements over the baseline across all scales. We observe the most significant added value in small scales (65–500 km), specifically in the offshore regions (both high and low variability), which we have already shown in the quantitative analysis in the Table~\ref{tab:relative_error_leadtime0_regions}. We suggest that the 4DVarNet forecast model is able to capture smaller dynamics due to its gradient regularization. Similarly, UNet also shows more improvements in the filtered maps, with a particular gain in ofsshore low variability regions, with a localized degradations along the equatorial band. In the contrast to the end-to-end neural forecasts, GloNet shows heterogeneous improvements wrt baseline GLO12 across the chosen metric, with a particular degradation of performances in coastal regions, and some improvements in higher latitudes. These spatial distributions of explained variance errors highlight that 4DVarNet provides the most consistent improvements globally, whereas UNet and GloNet show more heterogeneous performances. Furthermore, the more significant improvements at smaller scales underscores that the end-to-end neural forecasts can better exploit sparse observations to forecast dynamics at scales where traditional model-based forecast is less reliable.

\section{Conclusion and Discussion}\label{s: discussion}

This study introduces a deep learning-based framework for short-term ocean forecasting, leveraging end-to-end neural architecture, especially 4DVarNet schemes, in a novel forecasting setup. Trained on simulated nadir altimetry data, the model demonstrates state-of-the-art performance in predicting global SSH over a 7-day horizon. Integrated within the OceanBench benchmarking initiative \citep{johnson2023oceanbench, eloceanbench}, our approach provides a standardized and reproducible framework for neural ocean forecasting. These results open promising avenues for future improvements in operational oceanography using machine learning. Our findings demonstrate the relevance of end-to-end neural networks for short-term ocean forecasting. The ability of 4DVarNet to handle sparse observations and to produce accurate forecasts is of particular interest for operational contexts, where real-time data are often incomplete or noisy.

While we only rely on training strategies based on reanalysis datasets and simulated observation data, future work could explore the potential of training schemes using real observation datasets \citep{martin2024deep, dorffer2025observation}. In this context, the benefit of wide-swath altimetry data could be highly valuable \citep{ballarotta2025integrating, fouchet2025comparison}. Accounting for uncertainty quantification in the proposed end-to-end neural forecasts is also a key challenge. We aim to explore the combination of 4DVarNet schemes and generative approaches \citep{li2024generative} as recently explored for SLA mapping \citep{beauchamp2025neural}. 

\section*{Acknowledgments}

This work has been carried out as part of the Copernicus Marine Service OceanBench-STOF project. Copernicus Marine Service is implemented by Mercator Ocean in the framework of a delegation agreement with the European Union. This work was partially supported by the French ANR OceaniX (ANR-19-CHIA-0016). It also benefited from HPC and GPU resources provided by GENCI-IDRIS (Grant 2021-101030) and the CPER AIDA GPU cluster, supported by the Regional Council of Brittany, Brest Métropole, and Fonds Européen de DEveloppement Régional (FEDER).

\bibliography{bibliography}

\begin{thebibliography}{}

\bibitem[Aouni et~al., 2025]{eloceanbench}
Aouni, A.~E., Gaudel, Q., Johnson, J.~E., Charly, R., Sommer, J.~L., van
  Gennip, Fablet, R., Drevillon, M., DRILLET, Y., and Traon, P. Y.~L. (2025).
\newblock Oceanbench: A benchmark for data-driven global ocean forecasting
  systems.
\newblock In {\em The Thirty-ninth Annual Conference on Neural Information
  Processing Systems Datasets and Benchmarks Track}.

\bibitem[Ashish, 2017]{ashish2017attention}
Ashish, V. (2017).
\newblock Attention is all you need.
\newblock {\em Advances in neural information processing systems}, 30:I.

\bibitem[Ballarotta et~al., 2025]{ballarotta2025integrating}
Ballarotta, M., Ubelmann, C., Bellemin-Laponnaz, V., Le~Guillou, F., Meda, G.,
  Anadon, C., Laloue, A., Delepoulle, A., Faug{\`e}re, Y., Pujol, M.-I., et~al.
  (2025).
\newblock Integrating wide-swath altimetry data into level-4 multi-mission
  maps.
\newblock {\em Ocean Science}, 21(1):63--80.

\bibitem[Ballarotta et~al., 2019]{ballarotta2019resolutions}
Ballarotta, M., Ubelmann, C., Pujol, M.-I., Taburet, G., Fournier, F., Legeais,
  J.-F., Faug{\`e}re, Y., Delepoulle, A., Chelton, D., Dibarboure, G., et~al.
  (2019).
\newblock On the resolutions of ocean altimetry maps.
\newblock {\em Ocean science}, 15(4):1091--1109.

\bibitem[Barth et~al., 2020]{barth2020dincae}
Barth, A., Alvera-Azc{\'a}rate, A., Licer, M., and Beckers, J.-M. (2020).
\newblock Dincae 1.0: A convolutional neural network with error estimates to
  reconstruct sea surface temperature satellite observations.
\newblock {\em Geoscientific Model Development}, 13(3):1609--1622.

\bibitem[Beauchamp et~al., 2021]{beauchamp2021end}
Beauchamp, M., Amar, M.~M., Febvre, Q., and Fablet, R. (2021).
\newblock End-to-end learning of variational interpolation schemes for
  satellite-derived ssh data.
\newblock In {\em 2021 IEEE International Geoscience and Remote Sensing
  Symposium IGARSS}, pages 7418--7421. IEEE.

\bibitem[Beauchamp et~al., 2025]{beauchamp2025neural}
Beauchamp, M., Fablet, R., Benaichouche, S., Tandeo, P., Desassis, N., and
  Chapron, B. (2025).
\newblock Neural variational data assimilation with uncertainty quantification
  using spde priors.
\newblock {\em Artificial Intelligence for the Earth Systems}, 4(3):240060.

\bibitem[Bi et~al., 2023]{bi2023accurate}
Bi, K., Xie, L., Zhang, H., Chen, X., Gu, X., and Tian, Q. (2023).
\newblock Accurate medium-range global weather forecasting with 3d neural
  networks.
\newblock {\em Nature}, 619(7970):533--538.

\bibitem[Brasseur and Verron, 2006]{brasseur2006seek}
Brasseur, P. and Verron, J. (2006).
\newblock The seek filter method for data assimilation in oceanography: a
  synthesis.
\newblock {\em Ocean Dynamics}, 56(5):650--661.

\bibitem[Buizza et~al., 2018]{buizza2018development}
Buizza, R., Balmaseda, M.~A., Brown, A., English, S., Forbes, R., Geer, A.,
  Haiden, T., Leutbecher, M., Magnusson, L., Rodwell, M., et~al. (2018).
\newblock {\em The development and evaluation process followed at ECMWF to
  upgrade the Integrated Forecasting System (IFS)}.
\newblock European Centre for Medium Range Weather Forecasts.

\bibitem[Cui et~al., 2025]{cui2025forecasting}
Cui, Y., Wu, R., Zhang, X., Zhu, Z., Liu, B., Shi, J., Chen, J., Liu, H., Zhou,
  S., Su, L., et~al. (2025).
\newblock Forecasting the eddying ocean with a deep neural network.
\newblock {\em Nature Communications}, 16(1):2268.

\bibitem[Cummings and Smedstad, 2013]{cummings2013variational}
Cummings, J.~A. and Smedstad, O.~M. (2013).
\newblock Variational data assimilation for the global ocean.
\newblock In {\em Data assimilation for atmospheric, oceanic and hydrologic
  applications (Vol. II)}, pages 303--343. Springer.

\bibitem[Davidson et~al., 2009]{davidson2009applications}
Davidson, F.~J., Allen, A., Brassington, G.~B., Øyvind Breivik, Daniel, P.,
  Kamachi, M., Sato, S., King, B., Lefevre, F., Sutton, M., and Kaneko, H.
  (2009).
\newblock Applications of godae ocean current forecasts to search and rescue
  and ship routing.
\newblock {\em Oceanography}, 22(3):176--181.

\bibitem[Dorffer et~al., 2025]{dorffer2025observation}
Dorffer, C., Jourdin, F., Nguyen, T. T.~N., Devillers, R., Mouillot, D., and
  Fablet, R. (2025).
\newblock Observation-only deep learning for gappy satellite-derived ocean
  colour data using 4dvarnet.
\newblock {\em IEEE Transactions on Geoscience and Remote Sensing}.

\bibitem[Ducet et~al., 2000]{ducet2000global}
Ducet, N., Le~Traon, P.-Y., and Reverdin, G. (2000).
\newblock Global high-resolution mapping of ocean circulation from
  topex/poseidon and ers-1 and-2.
\newblock {\em Journal of Geophysical Research: Oceans}, 105(C8):19477--19498.

\bibitem[El~Aouni et~al., 2025]{el2025glonet}
El~Aouni, A., Gaudel, Q., Regnier, C., Van~Gennip, S., Le~Galloudec, O.,
  Drevillon, M., Drillet, Y., and Lellouche, J.-M. (2025).
\newblock Glonet: Mercator's end-to-end neural global ocean forecasting system.
\newblock {\em Journal of Geophysical Research: Machine Learning and
  Computation}, 2(3).

\bibitem[Fablet et~al., 2021]{fablet2021learning}
Fablet, R., Chapron, B., Drumetz, L., M{\'e}min, E., Pannekoucke, O., and
  Rousseau, F. (2021).
\newblock Learning variational data assimilation models and solvers.
\newblock {\em Journal of Advances in Modeling Earth Systems}, 13(10).

\bibitem[Fablet et~al., 2023]{fablet2023multimodal}
Fablet, R., Febvre, Q., and Chapron, B. (2023).
\newblock Multimodal 4dvarnets for the reconstruction of sea surface dynamics
  from sst-ssh synergies.
\newblock {\em IEEE Transactions on Geoscience and Remote Sensing}, 61:1--14.

\bibitem[Febvre et~al., 2024]{febvre2024training}
Febvre, Q., Le~Sommer, J., Ubelmann, C., and Fablet, R. (2024).
\newblock Training neural mapping schemes for satellite altimetry with
  simulation data.
\newblock {\em Journal of Advances in Modeling Earth Systems}, 16(7).

\bibitem[Febvre et~al., 2023]{febvre2023training}
Febvre, Q., Sommer, J.~L., Ubelmann, C., and Fablet, R. (2023).
\newblock Training neural mapping schemes for satellite altimetry with
  simulation data.
\newblock {\em arXiv preprint arXiv:2309.14350}.

\bibitem[Fennel et~al., 2022]{fennel2022ocean}
Fennel, K., Mattern, J.~P., Doney, S.~C., Bopp, L., Moore, A.~M., Wang, B., and
  Yu, L. (2022).
\newblock Ocean biogeochemical modelling.
\newblock {\em Nature Reviews Methods Primers}, 2(1):76.

\bibitem[Fouchet et~al., 2025]{fouchet2025comparison}
Fouchet, E., Benkiran, M., Le~Traon, P.-Y., and Remy, E. (2025).
\newblock Comparison of a global high-resolution ocean data assimilation system
  with swot observations.
\newblock {\em Frontiers in Marine Science}, 12:1563934.

\bibitem[Garcia et~al., 2025]{garcia2025orcast}
Garcia, P., Larroche, I., Pesnec, A., Bull, H., Archambault, T., Moschos, E.,
  Stegner, A., Charantonis, A., and B{\'e}r{\'e}ziat, D. (2025).
\newblock Orcast: Operational high-resolution current forecasts.
\newblock {\em Artificial Intelligence for the Earth Systems}.

\bibitem[Hobday et~al., 2016]{hobday2016hierarchical}
Hobday, A.~J., Alexander, L.~V., Perkins, S.~E., Smale, D.~A., Straub, S.~C.,
  Oliver, E.~C., Benthuysen, J.~A., Burrows, M.~T., Donat, M.~G., Feng, M.,
  et~al. (2016).
\newblock A hierarchical approach to defining marine heatwaves.
\newblock {\em Progress in oceanography}, 141:227--238.

\bibitem[Hoteit et~al., 2024]{hoteit2024improving}
Hoteit, I., Chassignet, E., and Bell, M. (2024).
\newblock Improving accuracy and providing uncertainty estimations: ensemble
  methodologies for ocean forecasting.
\newblock {\em State of the Planet Discussions}, 2024:1--11.

\bibitem[Johnson et~al., 2023]{johnson2023oceanbench}
Johnson, J.~E., Febvre, Q., Gorbunova, A., Metref, S., Ballarotta, M.,
  Le~Sommer, J., et~al. (2023).
\newblock Oceanbench: The sea surface height edition.
\newblock {\em Advances in Neural Information Processing Systems},
  36:78275--78295.

\bibitem[Lahoz and Schneider, 2014]{lahoz2014data}
Lahoz, W.~A. and Schneider, P. (2014).
\newblock Data assimilation: making sense of earth observation.
\newblock {\em Frontiers in Environmental Science}, 2:16.

\bibitem[Lam et~al., 2022]{lam2022graphcast}
Lam, R., Sanchez-Gonzalez, A., Willson, M., Wirnsberger, P., Fortunato, M.,
  Alet, F., Ravuri, S., Ewalds, T., Eaton-Rosen, Z., Hu, W., et~al. (2022).
\newblock Graphcast: Learning skillful medium-range global weather forecasting.
\newblock {\em arXiv preprint arXiv:2212.12794}.

\bibitem[Le~Traon et~al., 2017]{le2017use}
Le~Traon, P.-Y., Dibarboure, G., Jacobs, G., Martin, M., R{\'e}my, E., and
  Schiller, A. (2017).
\newblock Use of satellite altimetry for operational oceanography.
\newblock {\em Satellite altimetry over oceans and land surfaces}, pages
  581--608.

\bibitem[Le~Traon et~al., 2025]{le2025satellite}
Le~Traon, P.-Y., Dibarboure, G., Lellouche, J.-M., Pujol, M.-I., Benkiran, M.,
  Drevillon, M., Drillet, Y., Faug{\`e}re, Y., and Remy, E. (2025).
\newblock Satellite altimetry and operational oceanography: from jason-1 to
  swot.
\newblock {\em Ocean Science}, 21(4):1329--1347.

\bibitem[Le~Traon et~al., 2019]{le2019observation}
Le~Traon, P.~Y., Reppucci, A., Alvarez~Fanjul, E., Aouf, L., Behrens, A.,
  Belmonte, M., Bentamy, A., Bertino, L., Brando, V.~E., Kreiner, M.~B., et~al.
  (2019).
\newblock From observation to information and users: The copernicus marine
  service perspective.
\newblock {\em Frontiers in marine science}, 6:234.

\bibitem[Lellouche et~al., 2018]{lellouche2018mercator}
Lellouche, J.-M., Greiner, E., Le~Galloudec, O., Regnier, C., Benkiran, M.,
  Testut, C.-E., Bourdalle-Badie, R., Drevillon, M., Garric, G., and Drillet,
  Y. (2018).
\newblock Mercator ocean global high-resolution monitoring and forecasting
  system.
\newblock {\em New Frontiers in Operational Oceanography}, pages 563--592.

\bibitem[Lellouche et~al., 2023]{lellouche2023evolution}
Lellouche, J.-M., Greiner, E., Ruggiero, G., Bourdall{\'e}-Badie, R., Testut,
  C.-E., Le~Galloudec, O., Benkiran, M., and Garric, G. (2023).
\newblock Evolution of the copernicus marine service global ocean analysis and
  forecasting high-resolution system: potential benefit for a wide range of
  users.
\newblock {\em Proceeding Eurogoos}, pages 242--251.

\bibitem[Lellouche et~al., 2013]{lellouche2013evaluation}
Lellouche, J.-M., Le~Galloudec, O., Dr{\'e}villon, M., R{\'e}gnier, C.,
  Greiner, E., Garric, G., Ferry, N., Desportes, C., Testut, C.-E., Bricaud,
  C., et~al. (2013).
\newblock Evaluation of global monitoring and forecasting systems at mercator
  oc{\'e}an.
\newblock {\em Ocean Science}, 9(1):57--81.

\bibitem[Lermusiaux et~al., 2006]{lermusiaux2006quantifying}
Lermusiaux, P.~F., Chiu, C.-S., Gawarkiewicz, G.~G., Abbot, P., Robinson,
  A.~R., Miller, R.~N., Haley, P.~J., Leslie, W.~G., Majumdar, S.~J., Pang, A.,
  et~al. (2006).
\newblock Quantifying uncertainties in ocean predictions.

\bibitem[Li et~al., 2024]{li2024generative}
Li, L., Carver, R., Lopez-Gomez, I., Sha, F., and Anderson, J. (2024).
\newblock Generative emulation of weather forecast ensembles with diffusion
  models.
\newblock {\em Science Advances}, 10(13):eadk4489.

\bibitem[Liu et~al., 2023]{liu2023systematic}
Liu, G., Bracco, A., and Brajard, J. (2023).
\newblock Systematic bias correction in ocean mesoscale forecasting using
  machine learning.
\newblock {\em Journal of Advances in Modeling Earth Systems}, 15(11).

\bibitem[Madec et~al., 2015]{madec2015nemo}
Madec, G. et~al. (2015).
\newblock Nemo ocean engine.

\bibitem[Martin et~al., 2023]{martin2023synthesizing}
Martin, S.~A., Manucharyan, G.~E., and Klein, P. (2023).
\newblock Synthesizing sea surface temperature and satellite altimetry
  observations using deep learning improves the accuracy and resolution of
  gridded sea surface height anomalies.
\newblock {\em Journal of Advances in Modeling Earth Systems}, 15(5).

\bibitem[Martin et~al., 2024]{martin2024deep}
Martin, S.~A., Manucharyan, G.~E., and Klein, P. (2024).
\newblock Deep learning improves global satellite observations of ocean eddy
  dynamics.
\newblock {\em Geophysical Research Letters}, 51(17).

\bibitem[Ronneberger et~al., 2015]{ronneberger2015u}
Ronneberger, O., Fischer, P., and Brox, T. (2015).
\newblock U-net: Convolutional networks for biomedical image segmentation.
\newblock In {\em International Conference on Medical image computing and
  computer-assisted intervention}, pages 234--241. Springer.

\bibitem[Shi et~al., 2015]{shi2015convolutional}
Shi, X., Chen, Z., Wang, H., Yeung, D.-Y., Wong, W.-K., and Woo, W.-c. (2015).
\newblock Convolutional lstm network: A machine learning approach for
  precipitation nowcasting.
\newblock {\em Advances in neural information processing systems}, 28.

\bibitem[Taburet et~al., 2019]{taburet2019duacs}
Taburet, G., Sanchez-Roman, A., Ballarotta, M., Pujol, M.-I., Legeais, J.-F.,
  Fournier, F., Faugere, Y., and Dibarboure, G. (2019).
\newblock Duacs dt2018: 25 years of reprocessed sea level altimetry products.
\newblock {\em Ocean Science}, 15(5):1207--1224.

\bibitem[Wang et~al., 2024]{wang2024xihe}
Wang, X., Wang, R., Hu, N., Wang, P., Huo, P., Wang, G., Wang, H., Wang, S.,
  Zhu, J., and Xu, J. (2024).
\newblock Xihe: A data-driven model for global ocean eddy-resolving
  forecasting.
\newblock {\em arXiv preprint arXiv:2402.02995}.

\bibitem[Xiong et~al., 2023]{xiong2023ai}
Xiong, W., Xiang, Y., Wu, H., Zhou, S., Sun, Y., Ma, M., and Huang, X. (2023).
\newblock Ai-goms: Large ai-driven global ocean modeling system.
\newblock {\em arXiv preprint arXiv:2308.03152}.

\end{thebibliography}

\end{document}